# Fetching Strategy in the Startup Stage of p2p Live Streaming

Chunxi Li and Changjia Chen
School of Electronics and Information Engineering, Beijing Jiaotong University, 3 Shangyuancun, Haidian District, Beijing 100044, P. R. China
cxl@telecom.njtu.edu.cn and changjiachen@ sina.com).

*Abstract*—**A protocol named *Threshold Bipolar* (TB) is proposed as a fetching strategy at the startup stage of p2p live streaming systems. In this protocol, chunks are fetched consecutively from buffer head at the beginning. After the buffer is filled into a threshold, chunks at the buffer tail will be fetched first while keeping the contiguously filled part in the buffer above the threshold even when the buffer is drained at a playback rate. High download rate, small startup latency and natural strategy handover can be reached at the same time by this protocol. Important parameters in this protocol are identified. The buffer progress under this protocol is then expressed as piecewise lines specified by those parameters. Startup traces of peers measured from PPLive are studied to show the real performance of TB protocol in a real system. A simple design model of TB protocol is proposed to reveal important considerations in a practical design.**

## I. INTRODUCTION

P2p content distribution is one of most popular network applications today. The success of this technique lies on the ability to swap the capacity of all the individuals on an overlay network of peers. BitTorrent (BT) as the first prevalent p2p file sharing protocol designed by Bram Cohen [1] has appealed much interest of the research community. It has been proven a very effective mechanism for volume content distribution [2]. CoolStreaming [3] as a real prototype deployed on the real Internet demonstrates an application potential if we can distribute content not only in its volume but also distribute it lively. Research in [4] argued that a protocol is not necessarily applicable in live distribution only because it distributes volume content efficiently. Modifications of BitTorrent for live distribution then are studied in [4, 5, 6].

Many academic researches are stimulated, but the more important consequence of CoolStreaming is the blooming of p2p based IPTV applications. Among them, the most popular systems are PPLive, PPStream, QQLive et cetera. Both vast numbers of online peers they attracted and the measurement works published on them [7-12] indicates their success. In this paper, we will try to discussion the differences between the real protocol designs and previously published academic researches through a study on the fetching strategy of PPLive. PPLive is currently one of the most popular IPTV in China. Most measurement based studies on p2p streaming video are about this system [7-11]. In this paper, we will only study how a peer fetches chunk when he joins a p2p live streaming system. A protocol named as *Threshold Bipolar* (TB) is discussed as an applicable fetching strategy in the startup stage of a new peer. When a new peer joins the system, he must choose an initial offset for his buffer head, and then fetches chunks to fill his buffer. Measurement shows that, peers in a p2p live streaming system may buffer a large number of chunks. Hence, the chunks around the buffer head of this new peer have been saved in buffers of almost all other peers. Thus, the best strategy of a new peer is to fetch chunks from his buffer head for small startup latency. Since the chunk availability is not a constraint in this circumstance, a new peer can fill his buffer with a rate near his access rate, much higher than the video playback rate. However, other peers have to fill their buffers with the playback rate in consequence of chunk availability. The number of neighbors that have the chunk requested by a new peer will be decreased when the peer fetches more and more chunks. Once no one has the requested chunk, the new peer is better to change his fetching priority to fetch chunks from the buffer tail in assisting the distribution of the content. This outlines the basic fetching feature of TB protocol: buffer head first at the beginning and buffer tail first after. The detail of the TB protocol, its designing parameters and associated buffer progress of a new peer will be discussed in this paper. We will also verify that the TB protocol is likely adopted in PPLive. The buffer progress of a peer at startup stage in PPLive can be predicted by a model of TB protocol.

**Related Work**: Most measurement based studies on p2p live streaming are descriptive. They only report what they find and do not pay to much effort on explain why. The concept of buffer progress and related parameters such as buffer width, playable video and peer offset used in this paper are defined in [7-11]. The startup performance is addressed on the user perceptive sense [7]. Relative offset lag is noticed by [11] as the time difference between offset curves of two different peers. In this paper, we will study the startup process of new peers through each their buffer progress curves. By this way, the underneath protocol can be revealed instead of only statistics. Furthermore, the design considerations in practical environment can be discussed quantitatively. A service curve, which is defined as the chunk ID fed by tracker at each time, is introduced in this paper to form a common reference for all other progress curves. Simple mathematical formulas then can be easily applied in our analysis.

Fetching strategies are discussed in [1-6] on theoretical basis. The rarest first and the greedy (following the definition of [6]) as two extreme strategies have a long root at the BT like protocols. The mixed strategy, which uses both the rarest first and the greedy are proposed at [4] and [6]. Since they are proposed as a general strategy, the rarest first is chosen to have higher priority. The TB protocol in this paper is also mixed strategy at the startup stage. The considerations in this case are significantly different from the general application. In our work, the buffer progress curves as time sequences are derived theoretically and verified through real measured data. Our work is different from other works in following aspects:

- ✧ Working on large buffer width: The buffer widths used in PPLive and assumed in previous theoretical models are different essentially. For example, buffers less than 100 chunks are discussed in [6]. Theoretical analysis in [6] shows that, only in this region, different fetching strategies introduce significantly different buffer occupations. However, in PPLive more than 2000 chunks are saved by each peer in his buffer. It is doubtful to design a fetching protocol based only on the shape of buffer occupations in the tail of 10%.
- ✧ Startup and normal working stages: No working stages are considered in existing theoretical models. The scheme adopted by peers is the same no matter when a peer joins or after a peer goes stable. However, the environments and goals for a peer are different when he joins, and after he goes stable. When joining, a peer has an empty buffer and almost all chunks needed by this peer are saved on all other peers. How to fill his buffer head as soon as possible is the main consideration in this stage. After being stable, a peer has a buffer occupation similar to the occupations of all surrounding peers. Only in this stage, the buffer tail is the main consideration. In this paper, we will concentrate on the startup stage of a p2p streaming system.
- ✧ Deterministic description: Probabilistic method [3-6, 13,14] and fluid model [2,15-18] are used to describe the sharing environment in p2p content distribution. In this paper, we will model the buffer progress as piecewise lines. Each turn points in those piecewise lines are specified.

The organization of the paper is as follows. In Section 2, we will discuss possible models of initial fetching in p2p live streaming systems after a briefly outline on p2p live streaming systems. The protocol of the *Threshold Bipolar* (TB) is emphasized in this section. The initial fetching strategy in PPLive is addressed in section 3. A simple TB model is established which matches the buffer progress in PPLive fairly well. Design considerations are studied in this section as well. Section 4 concludes the paper.

## II. STARTUP PROCESS AND INITIAL CHUNK FETCH

### 2.1 A Brief on p2p Live Streaming System

A p2p live streaming system uses few servers (named as tracker) to support large number of audiences (named as peer). In a typical p2p live streaming system, the video data is divided to chunks identified by continuously assigned sequence numbers. The sequence number is also called the chunk ID in many papers. Content server (or Seeder) of the system injects chunks one by one into the system (actually it is selected or asked by peers). Every peer in the system has a buffer organized by chunks. A buffer message (BM) is an abstract description of this buffer. BM consists of an *offset*, which is the ID of the chunk at the buffer head, and a sequence of $\{0,1\}$. In other words, a bit map of a buffer is used to show which chunks are stored in this buffer. A value of 1 (0) at the $i^{th}$ position indicates that the chunk with an ID *offset* + $i$−1 has been (has not been) stored in that buffer.

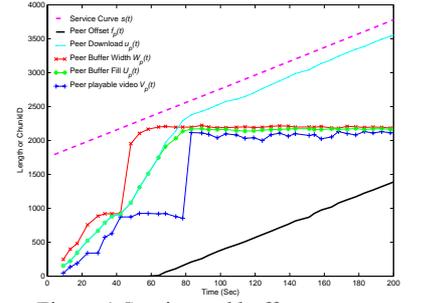

Figure 1 Service and buffer progress curves of a peer in PPLive

The startup process when peer joins a p2p live streaming system is studied in this paper. We will frequently use the *host* to name a new join peer. The name *peer* will mostly be used to address other peers the *host* connected. When a host joins, he first registers himself to a tracker. A list of current online peers is returned from the tracker after this registration. The host will try to connect peers in the list. Peer will send his current buffer message to the host once he is connected. The host then will choose a chunk ID in the streaming as a *start point* of his buffer to fetch chunks above this ID from other peers. This start point is also named as the *initial offset* in this paper. After certain time, the host will drain his buffer with the playback rate. In this paper, we will study how a host fetches chunks after the initial offset is determined.

### 2.2 Simple Model for Peer Buffer Progress

In this subsection, we will discuss a simple model on the peer buffer progress in a live streaming system. In this model, we will use $s(t)$ to indicate the *service curve* of the media stream. Where $s(t)$ is the largest chunk ID in the system at time $t$. For simplicity, we assume the media stream is CBR with a constant playback rate $r$. Peer drains his buffer with this rate [19]. The chunk ID in the buffer head and buffer tail of a peer $p$ at time $t$ are denoted by $f_p(t)$ and $\xi_p(t)$ respectively. We will name $f_p(t)$ as the *offset curve* and $\xi_p(t)$ as the *scope curve* of this peer respectively. Under the CBR assumption, the difference between the service and the offset curves of a given peer is independent to time. We will name the difference $L_p=s(t)-f_p(t)$ as the *offset lag* of peer $p$. In other sides, the *buffer width* $W_p(t)=\xi_p(t)-f_p(t)$ is fluctuated with time. Furthermore, we assume the tracker (or seeder) buffers the media contents with a width $W_{tk}$. Thus the tracker has an offset curve $f_{tk}(t)=s(t)-W_{tk}$. Except the buffer head and buffer tail, buffer occupations are also important in description of the buffer progress. Infinite patterns may be there in buffer occupations. To avoid too much detail in the description, we will only choose two parameters as the *playable video* $V_p(t)$ and *buffer fill* $U_p(t)$ in our model. The buffer fill $U_p(t)$ is the number of chunks in the buffer at time $t$. The playable video $V_p(t)$ is defined in [7] as the number of contiguous chunks in the buffer map, beginning from the offset. Correspondingly, the download curve of a peer $u_p(t)=U_p(t)+f_p(t)$ is the number of chunks the peer ever fetched. The instant download rate $r_p(t)$ of a peer is the derivative of download curve $r_p(t)=du_p(t)/dt$.

The service curve and peer buffer progress in a real system is shown in figure 1. In this paper we will call $W_p(t)$, $U_p(t)$ and $V_p(t)$ the buffer progress, and $\xi_p(t)$, $u_p(t)=U_p(t)+f_p(t)$ and $v_p(t)=V_p(t)+f_p(t)$ the peer progress of peer $p$. While the buffer progress indicates the relative chunk position in the buffer, the peer progress indicates the absolute value of chunk ID in the global sense.

*2.3 Protocol of Threshold Bipolar for Initial Chunk Fetch*

All published works to date only design fetching strategy for peers in an equilibrium environment. Some of them, for example [6], then use such designed protocol to evaluate the performance of peer when his joining. Many important aspects in sharing environment when a peer joins may be neglected in this way. For example, it is shown by [15-18] that, there is a phase-transition point $C(t)$ for each time $t$ in a p2p live distribution environment, any chunks with ID less than point $C(t)$ will be fetched readily. If a system adopts large peer buffer width like PPLive, most chunks buffered by any already online peer are readily fetched by a new peer. Form figure 1 one could also easily check that, the download rate $r_p(t)$ can be much higher than the video playback rate $r$ ($r_p(t) \approx 2.8r$ in this case) when a new peer fetches chunk below the phase-transition point. Thus, the method to estimate the startup latency suggested by [6] does not fit to a system with large buffer width. Based on these intuition establishments, one may consider designing a fetching strategy for startup stage to capture advantages in this special environment. The *Threshold Bipolar* (TB) proposed in this section is such a protocol for initial chunk fetch. The threshold bipolar strategy is simple in its implementation and designing philosophy. For a new peer, when the volume of playable video is lower than a threshold, the urgent task is to fill it up for smoothing playback. If the volume of playable video is above the threshold, the most important job is to spread the chunks from the source while keeping the volume of playable video above the threshold. This simple strategy is proven reliable and satisfactory by the practice in a largely deployed public network. Pseudo codes to implement TB protocol are listed in table 1.

There are two while loops, one is for the case when the playable video is larger than the threshold, and the other is for the case when the playable video is less than the threshold. If the playable video is less or equal to the threshold, a host (a peer who joins newly) will build his playable video step by step until the threshold is reached. If the playable video is larger than the threshold, a host will use two parallel threads to fetch chunks. FetchMax thread will fetch the maximum ID chunk among chunks that have not been fetched by the host. The FetchS thread will fetch chunk from the tracker. These codes provided here is only for understanding but not for implementation. The neighbor environment must be considered in a real implementation. For a given host $h$, let $X_h$ be those positions that have zeros in the BM of the host and ones in the BMs of at least one of his neighbors. Then instead of fetching chunk $f_h+V_h+1$ when playable video is lower than the threshold, a host will fetch the chunk $\min\{f_h+V_h+X_h\}$. Similar in FetchMax, a host will fetch chunk $\max\{f_h+X_h\}$.

*2.4 TB for Initial Chunk Fetch in a Real System*

In this section, we will use buffer progress curves in figure 1 to show that PPLive adopts a TB like protocol in its initial chunk fetch. We also use those curves to explain what typical buffer progress is looked like when TB protocol is applied. In figure 1, both buffer width curve $W_p(t)$ (red line with marker '$x$') and playable video curve $V_p(t)$ (blue line with marker '+') has a turnover point ($\tau_{sch}$, $C_{sch}$) at time $\tau_{sch}$ around 40 (seconds) and buffer position $C_{sch}$ around 900 (chunks). This point is corresponding to the event that the playable video hits the threshold at the first time. Before this time, only the first while loop in table 1 is executed, peer fetches chunks sequentially from smaller ID to larger ID. This can be checked from the closeness of buffer width $W_p(t)$, playable video $V_p(t)$ and buffer fill $U_p(t)$ (green line with marker '*'). Ideally, those three curves should be coincided in TB protocol. However, in real environment, a peer is hardly to successfully fetch chunk one by one sequentially in consequence of the next chunk may not be in buffers of all of his neighbors, or the fetching request is rejected by the neighbor (since this neighbor is busy in serving to other peer). There is a big jump of the buffer width $W_p(t)$ around $\tau_{sch}$. It is an evidence of the change in the protocol path. Therefore, we will name $\tau_{sch}$ as the *scheduling turnover time*. After the scheduling turnover time, the second while loop will be activated. The most advanced known chunk will be fetched. Thus, the buffer width will be enlarged suddenly just after the scheduling turnover time $\tau_{sch}$. In fact, second while loops is not the only codes executed at this stage. Once peer start to drain his buffer, the playable video will fall below the threshold. The first while loop will be activated to fetch the least advanced known chunk to fill up the playable video to the threshold. This can also be checked from figure 1. The offset $f_p(t)$ (thick black line) is only flat in the time interval [0, 65] since the buffer is drained after 65 seconds. However, The playable video $V_p(t)$ is nearly flat in the time interval [40, 80]. Least advanced chunks must be fetched in the time interval [65, 80] to keep the playable video just on the threshold. We will name the time when buffer starts to be drained as the *offset initial time* $\tau_{off}$. Buffer draining can also be checked from the difference of buffer fill curve $U_p(t)$ and download curve $u_p(t)$ (cyan line with marker '.'). As we have motioned before, the slope $u_p(t)$ is the peer download rate. Similarly, the slope $R_p(t)=dU_p(t)/dt$ is the buffer filling rate, which is equal to the difference of download rate $r_p(t)$ and the draining rate: $R_p(t)=r_p(t)-r$. Thus, those two curves will be coincided before $\tau_{off}$, and have different slope after $\tau_{off}$. Now let's look at buffer

TABLE 1. THRESHOLD BIPOLAR SCHEDULING

**Functions and Operations**:
**Fetch**($x$): request chunk with ID $x$
**Dec**($f_p$): {
 Drop the chunk with the ID $f_p$ form the buffer
 $f_p=f_p+1$; $V_p=V_p-1$;Right shift BM}
**WriteBM**($x$): {
 If ($x-f_p\leq$|BM|)
  Write the bit $x-f_p$ to 1;
 Else
  Expand BM to |BM|=$x-f_p$;
  Write the last bit of BM to 1;
 EndIf}
**FetchMax**: {
 *Fetch* ($f_p$+*position of last 0 in BM*);
 If(succeed) **WriteBM** (s)} EndIf
 }
**FetchS**: {
 Query tracker for $s$
 If(succeed ^ ($s>f_p$+|BM|)) *Fetch*(s)
  If(succeed) **WriteBM** (s)} EndIf
 EndIf}

**Main** {
 While ($V_p \leq Threshold$)
  Fetch ($f_p+V_p$ +1);
  If (succeed) **WriteBM**($f_p+V_p$ +1);
  endIf
 EndWhile

 While ($V_p > Threshold$)
  @ **FetchMax**
  @ **FetchS**:
 EndWhile
*EndMain*}

@: Can be parallel threads

progress curves at the time around 80 seconds. All curves of $W_p(t)$, $V_p(t)$ and $U_p(t)$ are converged to their stable state after this time. Thus, we will name this time as the *convergence time* $\tau_{cvg}$. A large jump in playable video $V_p(t)$ at convergence time $\tau_{cvg}$ indicates that the least advanced chunk does be fetched the last. It proves that the advanced chunk is fetched first in the time interval [$\tau_{sch}$, $\tau_{cvg}$] in an opposite direction.

In summaries, if we assume both the video playback rate $r$ and the peer download rate $r_p$ are invariant with time, then peer buffer progress curves in a TB protocol will be piecewise lines with turnover points or/and jumping points at the scheduling turnover time $\tau_{sch}$, the offset initial time $\tau_{off}$ and the convergence time $\tau_{cvg}$. All curves should be coincident with an increase rate $r_p$ in the time interval [$0,\tau_{sch}$]. The buffer width $W_p(t)$ has a jump around the scheduling turnover time $\tau_{sch}$ and quickly saturated after this time. The playable video curve $V_p(t)$ should be nearly horizontal at the time interval [$\tau_{sch},\tau_{cvg}$] and then saturated after a jump at the convergence time $\tau_{cvg}$. The buffer fill $U_p(t)$ is a piecewise line with a slope $r_p$ at the time interval [$0,\tau_{off}$] and a slope $r_p-r$ at the time interval [$\tau_{off},\tau_{cvg}$]. All curves are horizontal lines after the convergence time $\tau_{cvg}$.

Figure 2 shows buffer progress curves of six peers measured from PPLive. Peer 1 to 4 have a playback rate around 10 *chunks/s* and peer 5 and 6 have a playback rate around 6 *chunks/s*. Peer 1,4 and 6 have a scheduling turnover time $\tau_{sch}$, which is less than the offset initial time $\tau_{off}$ ($\tau_{sch}<\tau_{off}$). In opposite, Peer 2 and 3 have a scheduling turnover time $\tau_{sch}$, which is larger than the offset initial time $\tau_{off}$ ($\tau_{sch}>\tau_{off}$). Furthermore, peer 4 and 6 give us cases that buffer progress curves are converged before peer drains his buffer ($\tau_{cvg}<\tau_{off}$). Finally, peer 3 shows a peer with very poor startup environment. Except peer 3, the buffer progress of all other peers can be interpreted by TB protocol easily. Peer 3 has trouble in fetching chunks from his neighbors, but still his buffer progress follows the TB protocol roughly. The horizontal segment in the playable video and jumps at both playable video and buffer width can still be seen in the buffer progress of this peer. Therefore, we can conclude that TB protocol is likely to be applied in PPLive.

*2.5 Analytical model for Peer Progress under TB protocol*

Above discussions suggest that peer progress curves under TB protocol can be approximated by a bunch of piecewise lines. In this subsection, we will show that six parameters are critical in determine those piecewise lines for all buffer progress curves. Those six parameters are the video playback rate $r$, the peer download rate $r_p$, the threshold $C_{sch}$, the offset initial time $\tau_{off}$, the initial offset value $\theta_p$ and the offset lag $W^*$. The video playback rate $r$ is not a system design parameter since it will change program to program. The peer download rate $r_p$ is also not a system design parameter since it changes with different peers. Only the threshold $C_{sch}$, the offset initial time $\tau_{off}$, the initial offset $\theta_p$ and the offset lag $W^*$ can be designed in a TB protocol. In this section, we will discuses how to approximate buffer progress after those six parameters have been known. The design problem will be discussed in later sections. In following discussions, we will assume that the video playback rate $r$ is a constant. This assumption is reasonable since most p2p live streaming systems serve CBR programs. Furthermore, the startup stage is short (in an order of minutes), to assume a constant rate in this short time interval will easily be acceptable. If the video playback rate is constant, the offset curve of all online peers in our model are assumed to be $f(t)=rt$ except for the new peer. The offset $f_p(t)$ of a new peer $p$ is lineally progressed with playback rate $r$ after the offset initial time $\tau_{off}$.

$$f_p(t)=\theta_p+r(t-\tau_{off})^+. \qquad (1)$$

Where $(x)^+$ equals $x$ if $x>0$ and otherwise the value 0. No any theoretical result declares a constant peer download rate at the startup stage, but we still assume the peer download rate $r_p$ is a constant in this paper. We do so because:

1. Our measurements show that the peer download rate in startup stage is nearly constant.
2. As we have mentioned that theoretical results have proven the existence of a phase-transition point. Below this point, peer can download chunks readily near his access rate. We will show that this phase-transition point is very close to the situation region through measurement in later section. Therefore, to approximate a constant download rate at this stage is reasonable some how.
3. It is easy in mathematical treatment.

Now let's find the scheduling turnover time $\tau_{sch}$ and the convergence time $\tau_{cvg}$ after those six parameters of $r$, $r_p$, $C_{sch}$, $\tau_{off}$, $\theta_p$ and $W^*$ are given. It is easier to work at the peer progress curves instead of buffer progress curves, so we will look at peer progress first. Figure 3(a) show peer progress curves in a simple design model. There are three lines with a same slope of playback rate $r$ ($r=1$ in the figure, scaled according to PPLive). The line in top is the service curve

$$s(t)=rt+W^*. \qquad (2)$$

For simplicity, we assume the peer scope curve is identical to the service curve and peer has a saturation width $W^*$. The line at bottom is the offset curve $f_p(t)$ expressed by equation (1). The middle line is the turnover threshold curve

$$c_{sch}(t)=f_p(t)+C_{sch}. \qquad (3)$$

Those three lines are also drawn in figure 3(b-d). The line (in

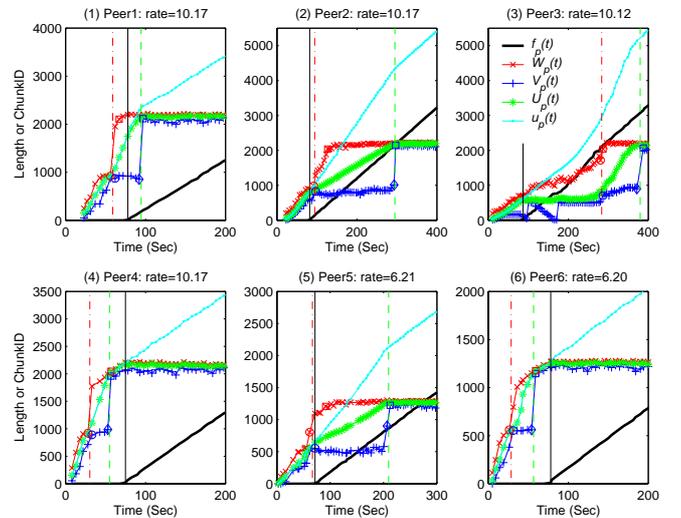

Figure 2. Buffer progress of peers in PPLive

green) which cross all of those three lines is the peer download curve

$$u_p(t) = r_p t + \theta_p. \quad (4)$$

By definition, the scheduling turnover time $\tau_{sch}$ is the first time when the threshold curve $c_{sch}(t)$ is reached by the download curve $u_p(t)$:

$$u_p(\tau_{sch}) = c_{sch}(\tau_{sch}) \Rightarrow C_{sch} = r_p \tau_{sch} - r(\tau_{sch} - \tau_{off})^+. \quad (5)$$

Solve equation (5) we have:

$$\tau_{sch} = \begin{cases} \dfrac{C_{sch}}{r_p}, & \text{if } C_{sch} \leq r_p \tau_{off} \\ \dfrac{C_{sch} - r\tau_{off}}{r_p - r}, & \text{if } C_{sch} > r_p \tau_{off} \end{cases} \quad (6)$$

Next let's find the convergence time $\tau_{cvg}$. The convergence time $\tau_{cvg}$ is the first time when the service curve is reached by the download curve $u_p(t)$:

$$u_p(\tau_{cvg}) = s(\tau_{cvg}) \Rightarrow r_p \tau_{cvg} + \theta_p = r\tau_{cvg} + W^*. \quad (7)$$

Solve equation (7) we have:

$$\tau_{cvg} = (W^* - \theta_p)/(r_p - r). \quad (8)$$

For simplifying notation, for any given function $f(t)$, we will write $f(t; t_0, t_1)$ as the function with value $f(t)$ when $t_0 < t < t_1$ and 0 otherwise. The peer progress curves now can be written as

$$\xi_p(t) = u_p(t; 0, \tau_{sch}) + s(t; \tau_{sch}, \infty);$$
$$v_p(t) = u_p(t; 0, \tau_{sch}) + c_{sch}(t; \tau_{sch}, \tau_{cvg}) + s(t; \tau_{cvg}, \infty); \quad (9)$$

Drawing curves actually is little bit messy due to the discontinuity in the offset curve. We have three possible orders in the three time instances: $\tau_{off} < \tau_{sch} < \tau_{cvg}$, $\tau_{sch} < \tau_{off} < \tau_{cvg}$ and $\tau_{sch} < \tau_{cvg} < \tau_{off}$. Correspondingly, the peer download rate can be divided into three rate group $\Gamma_0(r) = \{r_p: r < r_p \leq C_{sch}/\tau_{off}\}$, $\Gamma_1(r) = \{r_p: C_{sch}/\tau_{off} < r_p \leq r + (W^* - \theta_p)/\tau_{off}\}$ and $\Gamma_2(r) = \{r_p: r_p > r + (W^* - \theta_p)/\tau_{off}\}$ for given playback rate $r$. The peer progress for those rate groups are drawn in figure 3 (b), (c), and (d) respectively.

### III. TB PROTOCOL IN PPLIVE

*3.1 A Brief on the PPLive*

PPLive is currently one of the most popular IPTV to date [7-11]. In this section, we will use this practical system to show how TB protocol is designed in real applications. Like all current p2p live streaming applications, PPLive is a proprietary system. That means no any documents are published and no any program codes are opened for this system. Reverse engineering has to be involved to guess what is going on in this system through extensive measurement.

In this section, we will name a new arriving peer as a host. Based on our measurement, a host will receive tracker reply 0.06 seconds (on average) later after he registers himself to a tracker. The tracker reply includes a list of peers and two offset fields: *TkOffMin* and *TkOffMax*. The host will try to connect peers in the peer list and receive reply from other peers 1.42s later on average. The host will choose a chunk ID $\theta$ as his buffer head based on those replies. After about 2.57s on average, the host will fetch chunks above $\theta$ from connected peers. The host will advertise his offset 5s later. We will name $\theta$ as the initial offset. About 70s later, the host starts to drain his buffer at video playback rate $r$. In previous section, we have taken the time when a host starts to fetch chunk as the time reference 0. However, in a real system, host joins the system at different time. In this section, we will name the time when a host $h$ starts to fetch chunk as the host up time $t_h$. In this case, we have to use two definitions to differentiate between the *offset initial time* and *offset setup time*. The *offset initial time* denoted by $t_{off}$ is defined as the time when a host starts to drain his buffer. The *offset setup time* denoted by $\tau_{off}$ is defined as the time interval from the time when a peer starts to fill his buffer to the time when host starts to drain the buffer. Thus, we have

$$t_{off} = t_h + \tau_{off}. \quad (10)$$

Correspondingly, we will denote by $t_{sch}$ and $t_{cvg}$ the scheduling turnover time and the convergence time of a host respectively. The time interval of the scheduling turnover and the convergence will be $\tau_{sch} = t_{sch} - t_h$ and $\tau_{cvg} = t_{cvg} - t_h$.

A peer in PPLive will contact with the tracker and receive *TkOffMin* and *TkOffMax* nearly periodically at discrete time instances $\{t_i\}$. The value of *TkOffMax*($t_i$) at these discrete time instances is a sampling of the service curve $s(t)$ defined in subsection 2.2. Similar, the value of *TkOffMin*($t_i$) is a sampling of the offset curve $f_{tk}(t)$ of the tracker at time $t_i$. The difference of them is the buffer width $W_{tk}(t_i) = TkOffMax(t_i) - TkOffMin(t_i)$ of the tracker. Furthermore, we have found that the buffer width of the tracker is the scaled current playback rate $r(t_i)$:

$$r(t_i) = W_{tk}(t_i)/120. \quad (11)$$

In other words, a tracker will buffer 2 minutes contents for each channel in PPLive.

*3.2 Design Parameters of TB Protocol in PPLive*

As we have mentioned before, only four parameters of the threshold $C_{sch}$, the offset setup time $\tau_{off}$, initial offset value $\theta_p$ and the offset lag $W^*$ could be designed in a TB protocol. In this subsection, we will try to find how those parameters are designed in PPLive. In the discussion of this section, we will use data collected from Apr. 2 to Jul. 15,2007 through our crawler.

*3.2.1 Offset setup time*

In measurements, our crawler keep to trace each peer in a given channel and records the buffer message returned by each peer with a time stamp to indicate the time this message is received. Thus we will have the offset value of $f_p(t)$ at discrete time instances $\{t_{p,i}\}$ for each peer $p$. Not every peer caught by our crawler is a host since many peers enter the system earlier

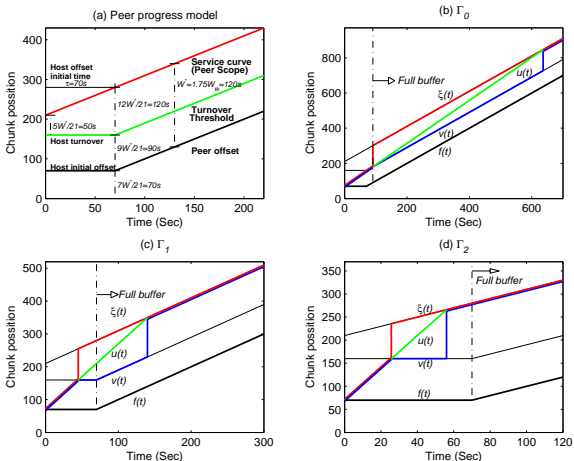

Figure 3. Distributions of offset setup time

Table 2. Mean and standard deviation of offset setup time

| Threshold of initial buffer fill | Num. of samples | Arithmetic Average | | Linear Interpolation | |
|---|---|---|---|---|---|
| | | Mean | Std | Mean | Std |
| 50 | 854 | 67.04 | 5.49 | 66.99 | 4.84 |
| 100 | 1909 | 66.56 | 5.32 | 66.46 | 5.07 |
| 500 | 3603 | 65.20 | 6.28 | 65.14 | 6.09 |

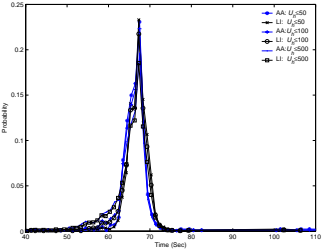
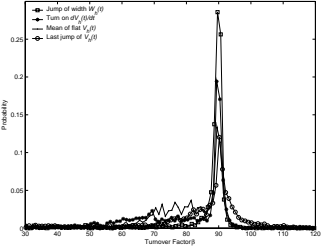

Figure 4. Distributions of offset setup time $T_{off}$

Figure 5. Distributions of turnover factor $\beta$

than our crawler does. A host should have a constant offset at his earlier records and then increased offsets in his later records. There are two problems in inferring the exact value of offset setup time $\tau_{off}$ from records of a given host. One is that, the reference point $T_{ad}$, when the time a host starts to advertise his offset, is often missed in the records. A crawler knows a host only through the peer list of the tracker. Hence, in most of the cases, a host already starts his process before our crawler queries him. Another problem is to find the exact time that a host start to drain his buffer because the time span between to consecutive records for some host may last for several tens seconds. Buffer fill $U_h(t)$ are used to overcome the first problem. We will only choose those hosts with initial buffer fills smaller than a certain given threshold, and then use the time stamp of the first record as the estimation of the advertising time $T_{ad}$ of this host. To check if the threshold in initial buffer fills will introduce biases, different threshold for initial buffer fills are compared. To overcome the second problem, we use two different methods to estimate the time when a host changes his offset. More precisely, let $t_1$ and $t_2$ be the earliest time pair that our crawler receives two consecutive reports with different offset values $f(t_1) \neq f(t_2)$ from a host. One method we have tried is based on a simple *arithmetic average* (AA) $t=(t_1+t_2)/2$. Another is based on *linear interpolation* (LI) $t=t_2-(f(t_2)-f(t_1))/r$. The offset initial time $T_{off}$ for a host then is calculated as $T_{off}=t-t_0$, where $t_0$ is the time our crawler receiving the first report from this host. The means and standard deviations of $T_{off}$ estimated by AA and LI with a different initial buffer fill threshold of 50, 100 and 500 chunks are listed in Table 2. Distributions of $T_{off}$ are also drawn in figure 4. Since all methods give out similar results that validates our methods in estimation of offset setup time. Based on this analysis we can get the conclusion that:

**Observation 1**: PPLive adopts a constant offset setup time $\tau_{off}$.

$$\tau_{off}=T_{off}+T_{ad}-T_{\theta}\approx 70 \text{ (seconds)}. \tag{12}$$

Where $T_{off}$, $T_{ad}$ and $T_{\theta}$ are the time of offset initial time, first advertising time and the time the host chooses his initial offset. The time $T_{off}$ can be measured through above discussed method. The first advertising time $T_{ad}$ is always a constant of 5s. We have not find a method to measure $T_{\theta}$ directly, but it can be approximated by the receiving time of the first reply from other peers. It is about 3.6 seconds by our measurement.

### 3.2.2 Turnover Threshold Factor

Even though the scheduling turnover can be observed in most host traces in our measurement, but it is not easy to find exact rule form the measurement data directly. Look at figure 1, we can see many ways to calculate the turnover threshold $C_{sch}$.

Table 3. Mean and spike of turnover factor $\beta$

| Method | Samples | Mean | Spike |
|---|---|---|---|
| Jump of width $W_p(t)$ | 1767 | 91.35 | 89.64 |
| Turn of $dV_p(t)/dt$ | 1524 | 79.27 | 89.38 |
| Mean of flat $V_p(t)$ | 1524 | 80.70 | 89.45 |
| Jump of pv $V_p(t)$ | 2501 | 77.65 | 90.18 |

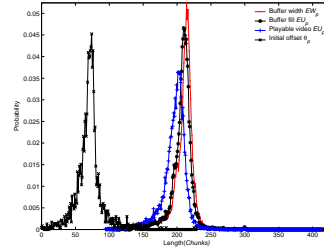
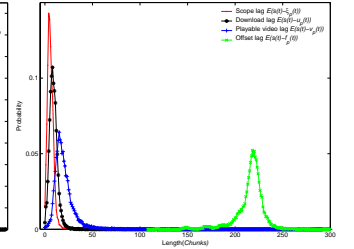

Figure 6. Distributions of average buffer progress

Figure 7. Distributions of average progress lag

✧ We can find the first jump in the buffer width $W_p(t)$ and let the threshold $C_{sch}$ equal the buffer width just before the jump.

✧ We can find the time when $dV_p(t)/dt$ changes at the first time and let the threshold $C_{sch}$ equal the playable video just after the change.

✧ We can let the threshold $C_{sch}$ equal the mean vale of playable video $V_p(t)$ on its flat part.

✧ We can find the jump in the playable video $V_p(t)$ and let the threshold $C_{sch}$ equal the playable video just before the jump.

We have used all four methods to estimate the turnover threshold $C_{sch}$. Instead of calculating the threshold directly, we try to find the threshold normalized by the playback rate $r$.

$$\beta=C_{sch}/r. \tag{13}$$

We will name $\beta$ as the turnover threshold factor. The probability distributions of turnover threshold factor $\beta$ calculated by above four methods are shown in figure 5. Some numerical characters are listed in table 3. All distributions have a spike at $\beta=90$. Based on this, we have following observation:

**Observation 2:** The turnover threshold factor $\beta$ in PPLive is a constant about 90.

### 3.2.3 The scheduling turnover and convergence time in PPLive

In this subsection, we will try to estimate the scheduling turnover time $\tau_{sch}$ and the convergence time $\tau_{cvg}$ in PPLive on a normalized basis. We will normalize all buffer progress and lags by the playback rate $r$. We also normalize the peer download rate by the playback rate $r$ as

$$\gamma_p=r_p/r. \tag{14}$$

By this way, the estimation will be independent to the playback rate. For example, the equation (3) for the scheduling turnover time $\tau_{sch}$ now reads

$$\tau_{sch} = \begin{cases} \beta\gamma_p^{-1}, & \text{if } \beta \leq \gamma_p \tau_{off} \\ (\beta - \tau_{off})(\gamma_p - 1)^{-1}, & \text{if } \beta > \gamma_p \tau_{off} \end{cases} \quad (15)$$

If we substitute the inferred value of $\beta$=90 and $\tau_{off}$=70 into (15), we have

$$\tau_{sch} = \begin{cases} 90\gamma_P^{-1}, & \text{if } \gamma_p \geq 9/7 \\ 20(\gamma_p - 1)^{-1}, & \text{if } \gamma_p < 9/7 \end{cases} \quad (16)$$

To find an estimation of the convergence time $\tau_{cvg}$, we need know the saturated buffer width of $W^*$. Furthermore, to find the buffer progress curves in startup stage, we need statistics of the saturation value in all buffer progress curves. So let's try to find these saturated values first.

For simplicity, we will use the same names and notations for a buffer progress curve before and after the normalization. Total 15,831 peer traces that lasts at least 5 min after saturation are found in our measurement. The distributions of average buffer width $W_p(t)$, buffer fill $U_p(t)$ and playable video $V_p(t)$ of the saturated segment of those traces are shown in figure 6. The figure 7 shows the distributions of the average progress lags of those peer traces. The numerical characters of those distributions are also listed in Table 4. In this computation, we first take the average for each peer trace, then the distribution, average and stand deviation of those peer average is calculated, shown in figure 6-7 and listed in table 4.

The initial offset value $\theta_p$ shown in figure 6 and listed in table 4 is measured through a client monitoring. Its mean value is about 70. This is very important in the design of PPLive. We will discuss it in later section.

The download rate at startup stage can only be measured through host peers. The distributions of normalized peer download rate $\gamma_p = r_p/r$ is shown in figure 8. The mean and stand deviation of those distributions are listed in table 5. Two methods are used in calculate the download rate of a peer. In *E2E* method we simply calculate the download rate $r_p=(u_p(t_1)-u_p(t_0))/(t_1-t_0)$, where $t_0$ and $t_1$ is the time of first report and the last report before saturation of a peer trace. In *Seg* method, we first calculate the instant rates of each consecutive reports, and then take the average of those instant rates as the download rate $r_p$ for each peer. To check if the distribution of peer download rate is related to actual playback rate, we also break peer traces into two different playback rate groups. One group has a playback rate around 10 *chunks/s* and the other 6 *chunks/s*. All of those distributions and statistics for different methods and different groups are similar. It means that peer download rate is relative smooth in startup stage. Perhaps, the similarity in different playback rate groups is more interested. As we have mentioned before, since the TB protocol fetches chunks when they are most available, a peer could have a download rate much larger than the actual playback rate. If we assume the access rates for peers are similar, then we would observe significantly higher normalized download rate at 6 *chunks/s* group than that of 10 *chunks/s* group. However, the rate enlargement can only be observed at small download rates (left part of the figure 8) but not at large download rates (right part of the figure 8). This may indicates that, the download rate is not totally dominated by the access bandwidth of a peer, otherwise the normalized download rate for 6 *chunk/s* group should be significantly larger than that of 10 *chunks/s* group. It shows certain aspects in network organization may be here to constraint the download rate below the actual bandwidth of a peer. Based on above measurement, we can use the average saturated buffer width $W^*$=210 to estimate the convergence time $\tau_{cvg}$. In this case, we have

$$\tau_{cvg} = 140(\gamma_p - 1)^{-1}. \quad (17)$$

Table 4. Average Buffer Progress and Progress Lags

|  | Samples | Mean | Std |
|---|---|---|---|
| Buffer width $EW_p$ | 15831 | 210.27 | 16.12 |
| Buffer fill $EU_p$ | 15831 | 20702 | 16.26 |
| Playable video $EV_p$ | 15831 | 195.98 | 18.21 |
| Initial offset $\theta_p$ | 2320 | 70.84 | 19.44 |
| Scope lag $E\{s(t)-\xi_p(t)\}$ | 15831 | 5.93 | 5.49 |
| Download lag $E\{s(t)-u_p(t)\}$ | 15831 | 9.11 | 6.08 |
| Playable lag $E\{s(t)-v_p(t)\}$ | 15831 | 20.13 | 11.80 |
| Offset lag $E\{s(t)-f_p(t)\}$ | 15831 | 216.10 | 17.25 |

### 3.2.4 Predicted buffer progress in PPLive

The time instances $\tau_{off}$, $\tau_{sch}$ and $\tau_{cvg}$ can only have three different orders: $\tau_{off}<\tau_{sch}<\tau_{cvg}$, $\tau_{sch}<\tau_{off}<\tau_{cvg}$ and $\tau_{sch}<\tau_{cvg}<\tau_{off}$. The buffer progress has three different patterns according to these orders. Thus, We can divide the download rate into three groups: $\Gamma_0=\{\gamma_p: 1<\gamma_p\leq 1.286\}$, $\Gamma_1=\{\gamma_p: 1.286<\gamma_p\leq 3\}$ and $\Gamma_2=\{\gamma_p: \gamma_p >3\}$. The buffer progress of these groups is depicted in figure 9 (b), (c) and (d) respectively. The probability that a peer belongs to each group is also listed in table 5.

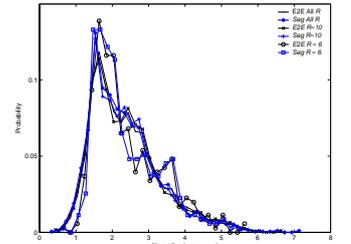

Figure 8. Distributions of download rates $\gamma_p$

The time instances in the group $\Gamma_0$ have an order of $\tau_{off}<\tau_{sch}<\tau_{cvg}$. Peer in this group faces a very poor startup environment. Table 5 shows that, very few peers belong to this group (less than 10%) in our measurement. Buffer progress in this group takes very long time to converge. The convergence time is in the range from infinite ($\gamma_p$=1, never converge) to 490s (about 8min). The time instances in the group $\Gamma_1$ have an order of $\tau_{sch}<\tau_{off}<\tau_{cvg}$. Most peers belong to this group (more than 70%). Hence this is the normal situation when peer startup in PPLive. The convergence time is in the range from 490s to 70s. Peers in the group $\Gamma_2$ have fast startup process. Buffer progress has converged before the peer drain his buffer. More than 20% peers belong to this group. Piecewise line representations of buffer progress in those different groups are listed in Table 6. Based on the table one can easily draw buffer progress when normalized download rate is given. If we apply the piecewise line model into peer traces in figure 2, we can find the predicted buffer progress in figure 10. Only groups $\Gamma_1$ (Peer 1, 2 and 3) and groups $\Gamma_2$ (Peer 4,5 and 6) are involved in those traces. Except peer 3, all other traces are fairly matched. In fact, peer 3 is working on a rather abnormal environment. At the beginning, the peer can only download chunks at a rate near the playback rate.

Table 5. Download rate and probability of groups

|  | Samp | Mean | Std | $\Gamma_0$ | $\Gamma_1$ | $\Gamma_2$ |
|---|---|---|---|---|---|---|
| E2E | 2037 | 2.390 | 0.973 | 0.074 | 0.703 | 0.224 |
| Seg | 2037 | 2.360 | 0.967 | 0.076 | 0.709 | 0.215 |
| E2E/10 | 1681 | 2.383 | 0.981 | 0.081 | 0.701 | 0.218 |
| Seg/10 | 1681 | 2.354 | 0.979 | 0.085 | 0.706 | 0.209 |
| E2E/6 | 353 | 2.422 | 0.932 | 0.034 | 0.714 | 0.252 |
| Seg/6 | 353 | 2.385 | 0.905 | 0.031 | 0.728 | 0.241 |

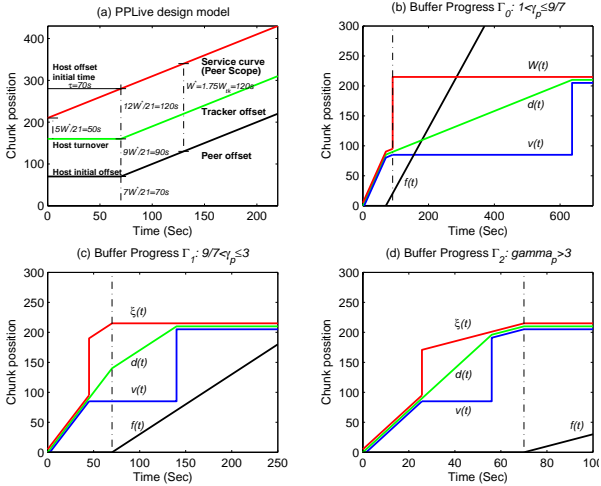

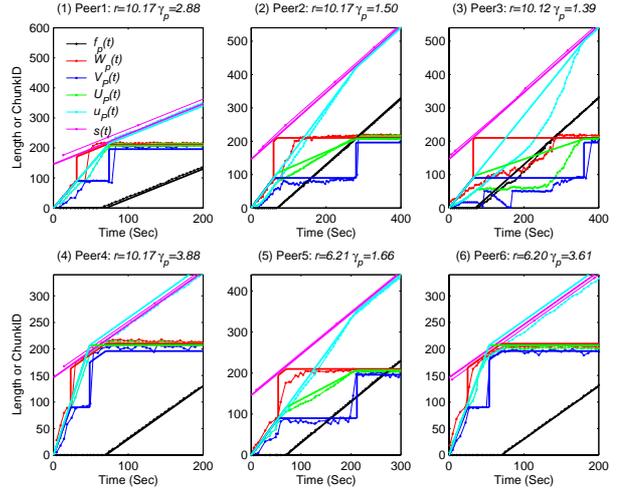

Figure 9. Design model and buffer progress in PPLive

Figure 10. Predicted buffer progress for traces in figure 2.

*3.3 Discussions on the Design Considerations in PPLive*

Figure 9(a) shows a design model for PPLive. We will assume the playback rate in this model is 1 *chunk/s*. If we normalize the chunk ID space by the playback rate in a practical system, we will get this design model. In this case, the buffer space can be interpreted as time duration. For example, a buffer width of 2000 chunks in a playback rate of 10 chunks/s will have a buffer width of 200 in our design model. The playback duration of those 2000chunks is happening to be 200s. In this simple model, we assume all peers except the new peer have identical offset curves $f(t)=t$ and same scope curves $\xi(t)$. Furthermore, we assume the peer scope curve is the same as service curve $s(t)=\xi(t)=t+W^*$. In this case when a peer arrives at time 0, he has to choose an initial offset value equals to the offset setup time $\theta_p=\tau_{off}$ in order to match his offset curve to that of other peers.

$\theta_p=\tau_{off}$.                                 (18)

Our measurements in figure 6 and table 4 validate this claim. Both offset setup time and initial offset value is about 70. In PPLive, the saturated buffer width $W^*$ is 210, so the initial offset is chosen to be 1/3 of the saturated buffer width:

$\theta_p=W^*/3$.              (19)

Since the saturated buffer width $W^*$ is also the playback duration of buffered chunks, we will also name it as buffer duration. In a p2p live streaming system, the offset setup time is roughly the startup latency. The buffer duration is the playback delay. Playback delay is not very sensitive for TV watching only. Therefore, people choose to use a relative long buffer duration to treat better playback continuity. In our simple model, the offset initial value will be fixed after the offset setup time has been chosen. It is natural to ask why people do not choose a smaller offset setup time for better startup latency. The answer is that, smaller offset setup time will mean smaller initial offset value. It also means tight requirement in the initial download rate since the remaining life in the initially fetched chunks will be shorter if the initial offset value is smaller.

For a given offset setup time $\tau_{off}$, the minimal download rate for fetching all initial $B$ chunks (chunks with ID form $\theta_p$ to $\theta_p+B-1$) is about $r_{min}=B/(\tau_{off}+B)$ since no one chunk can be survivor after $\tau_{off}+B$ seconds. If we short the offset setup time $\tau_{off}$ from 70s into 35s, the required minimal average download rates $r_{min}$ are compared in figure 11. For example, $r_{min}=0.5$ for first 70 chunks when $\tau_{off}=70s$. It is increased to $r_{min}=0.67$ when $\tau_{off}=35s$. In general, a peer has a poor initial download rate because he has very few neighbors at this stage. It should also be counted beside the small startup latency when design the offset setup time $\tau_{off}$.

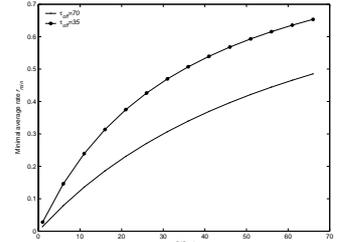

Figure 11. Minimal download rate for different $\tau_{off}$

There is a physical explanation for the turnover threshold $c_{sch}(t)$ in PPLive. In our normalized system, the turnover threshold curve defined by equation (3) can be expressed as

$c_{sch}(t)=\beta+\theta_p+(t-\tau_{off})^+ =\beta+\tau_{off}+(t-\tau_{off})^+$.  (20)

After $t\geq\tau_{off}$, we have

$c_{sch}(t)=t+\beta, t\geq\tau_{off}$.              (21)

It happens to be the tracker offset $f_{tk}(t)$ since

$f_{tk}(t)=t+W^*-120=t+90=t+\beta$.         (22)

Intuitively, chunks below the tracker offset are only saved by peers, but chunks above the tracker offset are saved by tracker as well as by peers. Peers may leaf but tracker will stay always. It is better if all peers in their buffers save those chunks that are no loge saved in tracker. In a normalized

Table 6. Table for piecewise line calculation

| $\Gamma_0=\{\gamma: 1<\gamma\leq 1.286\}$ | | | | $\Gamma_1=\{\gamma: 1.286<\gamma\leq 3\}$ | | | | $\Gamma_2=\{\gamma: \gamma>3\}$ | | | |
|---|---|---|---|---|---|---|---|---|---|---|---|
| $t$ | $V_p(t)$ | $W_p(t)$ | $U_p(t)$ | $t$ | $V_p(t)$ | $W_p(t)$ | $U_p(t)$ | $t$ | $V_p(t)$ | $W_p(t)$ | $U_p(t)$ |
| 0 | 0 | 0 | 0 | 0 | 0 | 0 | 0 | 0 | 0 | 0 | 0 |
| $\tau_o$ | $\gamma\tau_o$ | $\gamma\tau_o$ | $\gamma\tau_o$ | $\tau_s^-$ | $\gamma\tau_s$ | $\gamma\tau_s$ | $\gamma\tau_s$ | $\tau_s^-$ | $\gamma\tau_s$ | $\gamma\tau_s$ | $\gamma\tau_s$ |
| $\tau_s^-$ | $\gamma\tau_s-\tau_s+\tau_o$ | $\gamma\tau_s-\tau_s+\tau_o$ | $\gamma\tau_s-\tau_s+\tau_o$ | $\tau_s^+$ | $\beta$ | $W-\tau_o+\tau_s$ | $\gamma\tau_s$ | $\tau_s^+$ | $\beta$ | $W-\tau_o+\tau_s$ | $\gamma\tau_s$ |
| $\tau_s^+$ | $\beta$ | $W$ | $\gamma\tau_s-\tau_s+\tau_o$ | $\tau_o$ | $\beta$ | $W$ | $\gamma\tau_o$ | $\tau_c^-$ | $\beta$ | $W-\tau_o+\tau_c$ | $\gamma\tau_c$ |
| $\tau_c^-$ | $\beta$ | $W$ | $\gamma\tau_c-\tau_c+\tau_o$ | $\tau_c^-$ | $\beta$ | $W$ | $\gamma\tau_c-\tau_c+\tau_o$ | $\tau_c^+$ | $W-\tau_o+\tau_c$ | $W-\tau_o+\tau_c$ | $W-\tau_o+\tau_c$ |
| $\tau_c^+$ | $W$ | $W$ | $W$ | $\tau_c^+$ | $W$ | $W$ | $W$ | $\tau_o$ | $W$ | $W$ | $W$ |
| $\tau_m$ | $W$ | $W$ | $W$ | $\tau_m$ | $W$ | $W$ | $W$ | $\tau_m$ | $W$ | $W$ | $W$ |

$\beta=90$, $W=210$, $\tau_o=70$

PPLive system, the tracker offset is 120s and the peer buffer duration is 210s, so we will have a lower bound for $\beta$ as:

$$\beta\geq 90. \quad (23)$$

In other side, let $v_q(t)=V_q(t)+t$ be the peer progress of playable video of a neighbor peer $q$. All chunks with ID below $v_q(t)$ are already fetched by peer $q$ at time $t$, but the chunk with a ID of $v_q(t)+1$ has definitely not been fetched by peer $q$ at this time. If we set $c_{sch}(t)>v_q(t)$ and assume $q$ is the only neighbor currently for a new peer $p$, then after peer $p$ gets chunk $v_q(t)$, he has to idly wait peer $q$ to fetch the chunk $v_q(t)+1$ in TB protocol. If we design $c_{sch}(t)$ always less than $v_q(t)$, this kind of waste time will never happen. Thus we have an upper bound for $c_{sch}(t)$ as

$$c_{sch}(t)-v_q(t)<0, \text{ for } 0\leq t<\tau_{off}. \quad (24)$$

Substitute (20) into (24) we have

$$\beta<V_q(t)+t-\tau_{off}, \text{ for } 0\leq t<\tau_{off}. \quad (25)$$

If any possible download rates are considered, the right side of equation (25) has a minimal value $V_q(t)-\tau_{off}$. If we further assume the playable video curves for all peers are identically distributed process with a mean $V^*$ and stand deviation $\sigma_V$, then we have following design rule from (25)

$$\beta<V^*-\alpha\sigma_V-\tau_{off}, \text{ for } 0\leq t<\tau_{off}. \quad (26)$$

The purpose to introduce a coefficient $\alpha$ is to guarantee with large probability that the actual scheduling turnover of a new peer happens at a region below the playable video of his neighbor peers. Based on our measurement, $V^*$ is about 196 and $\sigma_V$ is about 18. For a threshold of 90, the value of $\alpha$ is 2.

## IV. CONCLUSION AND FUTURE WORK

Threshold Bipolar (TB) is proposed in this paper as a fetching strategy at the startup stage of p2p streaming systems. Its applicability is studied through measurement based reverse engineering to a popular IPTV application PPLive. Peer progress and buffer progress under this protocol are piecewise lines specified by few parameters. Those parameters are identified in this paper and their designed values in PPLive are inferred. A simple design model of TB protocol is introduced and design considerations are discussed in this paper.